\RequirePackage{amsmath}
\documentclass[12pt]{iopart}
\usepackage{iopams, mathptm}
\usepackage[english]{babel}
\usepackage[numbers,sort&compress]{natbib}
\usepackage[latin2]{inputenc}
\usepackage{bbold}
\usepackage{amsopn}
\usepackage{calrsfs}
\usepackage{subfigure}
\usepackage[T1]{fontenc}
\usepackage{graphicx}
\usepackage{color}
\usepackage{braket}
\usepackage{breqn}

\usepackage[unicode=true, breaklinks=false, pdfborder={0 0 1}, backref=false, colorlinks=true, linkcolor=blue, citecolor=blue]{hyperref}

\newcommand{\R}{\mathbb{R}}

\newcommand{\Sys}{\mathcal{S}}
\newcommand{\Env}{\mathcal{E}}
\newcommand{\oneA}{\mathcal{A}}

\newcommand{\iden}{\mathbb{1}}

\newcommand{\op}[1]{\mathbf{#1}}
\newcommand{\partTr}[2]{\text{Tr}_{#1}\left( #2 \right)}
\newcommand{\I}{\mathcal{I}}

\newcommand{\pk}{\mathfrak{I}}

\newcommand{\opH}{\op{H}}
\newcommand{\opS}{\op{S}}
\newcommand{\opB}{\op{B}}
\newcommand{\opP}{\op{P}}
\newcommand{\proj}[2]{\mathcal{P}_{#1}\left[ #2\right]}

\newcommand{\Hilbert}{\mathcal{H}}
\newcommand{\Banach}{\mathfrak{B}}
\newcommand{\M}{\op{M}}
\newcommand{\x}{\vec{x}}
\newcommand{\iso}[1]{\varphi\left[#1\right]}

\newcommand{\Acal}{\mathcal{A}}

\newcommand{\Ical}{\mathcal{I}}

\newcommand{\Ucal}{\mathcal{U}}

\newcommand{\Pcal}{\mathcal{P}}

\def\BraVert{\egroup\,\mid\,\bgroup}

\def\ketbra#1#2{\ket{#1\vphantom{#2}}\!\bra{#2\vphantom{#1}}}

\def\bra#1{\mathinner{\langle{#1}|}}

\def\ket#1{\mathinner{|{#1}\rangle}}

\def\braket#1{\mathinner{\langle{#1}\rangle}}

\begin{document}

\title{Non-Markovian quantum control as coherent stochastic trajectories}

\author{Fattah Sakuldee$^{1,a}$, Simon Milz$^{2,b}$, Felix A. Pollock$^{2,c}$, Kavan Modi$^{2,d}$}
\address{$^1$MU-NECTEC Collaborative Research Unit on Quantum Information\\ Department of Physics, Faculty of Science, Mahidol University, Bangkok 10400, Thailand}
\address{$^2$School Of Physics \& Astronomy, Monash University, Clayton, Victoria 3800, Australia}

\ead{$^a$fattah.sak@student.mahidol.ac.th, 
$^b$simon.milz@monash.edu,\\
\hspace{1cm} $^c$felix.pollock@monash.edu, $^d$kavan.modi@monash.edu}

\date{\today}
\begin{abstract}
We develop a notion of stochastic quantum trajectories. First, we construct a basis set of trajectories, called \emph{elementary trajectories}, and go on to show that any quantum dynamical process, including those that are non-Markovian, can be expressed as a linear combination of this set. We then show that the set of processes divide into two natural classes: those that can be expressed as convex mixture of elementary trajectories and those that cannot be. The former are shown to be \emph{entanglement breaking processes} (in each step), while the latter are dubbed \emph{coherent processes}. This division of processes is analogous to separable and entangled states. In the second half of the paper, we show, with an information theoretic game, that when a process is non-Markovian, coherent trajectories allow for decoupling from the environment while preserving arbitrary quantum information encoded into the system. We give explicit expressions for the temporal correlations (quantifying non-Markovianity) and show that, in general, there are more quantum correlations than classical ones. This shows that non-Markovian quantum processes are indeed fundamentally different from their classical counterparts. Furthermore, we demonstrate how coherent trajectories (with the aid of coherent control) could turn non-Markovianity into a resource. In the final section of the paper we explore this phenomenon in a geometric picture with a convenient set of basis trajectories.
\end{abstract}
\maketitle

\section{Introduction}

Almost a hundred years after the discovery of quantum mechanics, a great deal of effort is still going into formally quantifying quantum coherence~\cite{RevModPhys.89.041003}. Recent resource-theoretic approaches have been fruitful in highlighting the role of coherence in many quantum information applications~\cite{PhysRevLett.119.140402}. Quantum coherence is the minimal requirement for essential quantum phenomena, such as interference, nonlocality and contextuality. All of these phenomena can be seen as resulting from the existence of objects that cannot be explained in terms of statistical mixtures of those that are classically allowed. Nonetheless, quantum mechanics is a linear theory, and such objects can always be expressed as coherent combinations of the classical ones, e.g., superposition of states\footnote{The same phenomenon manifests itself when describing density matrices in terms of a fixed set of states, where there are always some that cannot be written as a convex mixture of that set.}. The superposition amplitudes are allowed to be complex numbers, which is a drastic departure from the statistical interpretation of classical theory.

In this article we aim to identify coherent dynamics. We use convexity as a guiding principle and define quantum dynamics that cannot be described as a convex mixture of classical events. To achieve this, we cast quantum dynamics in terms of stochastic trajectories. We first define a basis set of \emph{elementary trajectories} that can be interpreted as a sequence of classical events (measurement outcomes and preparations), before showing that any quantum dynamics can be described in terms of linear combinations of elementary trajectories. However, only a small set of dynamical processes can be written as convex mixture of these trajectories, and thus the remainder must be what we call \emph{coherent processes}. We show that these are a temporal analogue of entangled quantum states; in contrast, any process that is a convex mixture of elementary trajectories is shown to be a concatenation of entanglement breaking channels. Such processes are analogous to separable states, with the elementary trajectories corresponding to product states of a composite system. 

In the second half of the paper we focus on a simple class of open dynamics that lead to unital, but non-Markovian, dynamics for the system. Within this model we first construct the elementary trajectories and use them to investigate the interplay between purely irreversible action and reversible action. Dissipative unital dynamics has, as its defining feature, that the maximally mixed state as its fixed point. In other words, such dynamics destroy quantum information encoded onto a quantum system. However, It is well-known that if such dynamics are non-Markovian, then it is possible to dynamically decouple the system from the environment with the aid of appropriate control operations. Here, we cast this interesting phenomenon in terms of stochastic trajectories and with a concrete example we quantify non-Markovianity and recoverability of quantum information. Moreover, we argue that both non-Markovianity and coherent control are required to recover quantum information, and, via a scaling-unitary decomposition of the dynamics, substantiate this intuition with a geometrical interpretation. There has been a debate over what qualifies as a Markovian quantum process~ \cite{BreuerEA2016}. This question has been definitively answered recently in Ref.~\cite{pollock_operational_prl}, where operational conditions for quantum Markov processes are derived. In this article we follow this operational definition for Markov dynamics.

We begin in Sec.~\ref{sec:preliminary} by defining what we mean by classical and quantum stochastic trajectories, and how they relate to an underlying deterministic evolution. We then go on, in Sec.~\ref{sec:interfering}, to show how these can be coherently interfered to control the system in the presence of non-Markovianity, before presenting some examples related to decoupling in Sec.~\ref{sec:simple}. Finally, we conclude in Sec.~\ref{sec:conclusion}.

\section{Stochastic trajectories} \label{sec:preliminary}

Throughout this paper we consider finite dimensional systems (classical and quantum) that can be measured and prepared. The measurement outcomes are a finite discrete set, labeled as $x_{i_{\alpha}}$, where the subscripts $i$ and $\alpha$ signify the the corresponding measurement outcome and time step respectively. For example, $x_{i_{\alpha}}$ denotes, at time $t_{\alpha}$, the measurement outcome indexed by $i_{\alpha} \in \left\{1,\dots, d_{\alpha} \right\}$, where $d_{\alpha}$ is the number of perfectly distinguishable measurement outcomes, \textit{i.e.}, the dimension of the state space of the system. To simplify notation, we will choose equidistant time steps $t = t_{\alpha+1} - t_{\alpha}$ throughout this paper. After the measurement, the system can be reprepared, and the prepared states are denoted by $y_{j_{\alpha}}$, where the subscript labels one of a set of possible repreparations at time $t_{\alpha}$. In general, the set of outcomes at different times could be different; however, for simplicity of notation we will assume here that $d_{\alpha}=d$ does not depend on the time step.

\subsection{Classical stochastic trajectories} \label{sec:classical}

In classical statistical mechanics, the time evolution of a system can be described in terms of trajectories, which could, in principle, be observed by someone with access to the corresponding degrees of freedom. For example, an experimenter could measure the position of a particle undergoing Brownian motion at fixed times $t_N,\dots,t_1$. The sequence $x_{i_N},\dots,x_{i_1}$ of respective measurement outcomes defines a discrete trajectory of the particle, and the process of Brownian motion is described entirely once the joint probabilities $p(x_{i_N}, \dots ,x_{i_1})$ for every possible trajectory the particle can take are known. While the existence of an underlying continuous process, under minor assumptions, is guaranteed by the Kolmogorov extension theorem~\cite{kolmogorov_foundations_1956, feller_introduction_1968, breuer_theory_2007, tao_introduction_2011}, we will focus on the discrete case in this article for convenience of notation. We emphasize that the above scenario encompasses \textit{all} possible classically allowed dynamics.

Besides passively measuring the position of the particle, the experimenter could actively \textit{intervene}; upon measuring the particle at position $x_{i_{\alpha}}$, they could reset it at another position $y_{j_{\alpha}}$ before letting it continue to evolve\footnote{In principle, the number of possible repreparations could differ from the number of measurement outcomes. For compact notation we restrict our description to the case where these two numbers coincide.}. For such experimental scenarios, a trajectory is given by a sequence  $(x_{i_N};y_{j_{N-1}},x_{i_{N-1}}; \dots;y_{j_1},x_{i_1})$. The resulting process would be described by the joint probability distribution $p(x_{i_N};y_{j_{N-1}},x_{i_{N-1}},\dots; y_{j_1},x_{i_1}) := p(x_{i_N}, \dots,x_{i_1}|y_{j_{N-1}}, \dots,y_{j_1})$, \textit{i.e.}, the probability to find the particle at positions $x_{i_N},\dots,x_{i_1}$ given that it was reprepared at the positions $y_{j_{N-1}}, \dots,y_{j_1}$. Each of these probability distributions fully characterizes a \textit{trajectory} of the system~\cite{pearl_causality_2009,milz_kolmogorov_2017}.

In anticipation of our main subject matter, we adopt the language of quantum mechanics to discuss classical stochastic processes with interventions. First, consider the action of a measurement at $t_{\alpha}$ on a state $\rho$ that yields the outcome $x_{i_{\alpha}}$ followed by a repreparation of $y_{j_{\alpha}}$. It can be written as
\begin{gather}
\label{eqn::MeasRepClass}
\Pcal_{j_{\alpha} i_{\alpha}}[\rho] = \bra{x_{i_{\alpha}}}\rho\ket{x_{i_{\alpha}}} \ketbra{y_{j_{\alpha}}}{y_{j_{\alpha}}} = \mathrm{Tr}\left(\ketbra{x_{i_{\alpha}}}{x_{i_{\alpha}}}\rho\right) \ketbra{y_{j_{\alpha}}}{y_{j_{\alpha}}},
\end{gather}
where $\ket{x_{i_{\alpha}}}$ ($\ket{y_{j_{\alpha}}}$) is the definite state vector corresponding to the measurement outcome (repreparation) $x_{i_{\alpha}}$ ($y_{i_{\alpha}}$) and $\rho$ is a classical state, \textit{i.e.}, it is diagonal in the basis $\{\ket{x_{i_{\alpha}}}\}$. Clearly, the maps $\Pcal_{j_{\alpha} i_{\alpha}}$ are not trace preserving. However, they do form a convex-linear basis so that the action of any \textit{stochastic map} can be expressed as
\begin{gather}
\label{eqn:stochPrep}
P_{\alpha}[\rho] = \sum_{i_{\alpha} j_{\alpha}} \mu_{j_{\alpha}|i_{\alpha}} 
\bra{x_{i_{\alpha}}}\rho_{\alpha} \ket{x_{i_{\alpha}}} \ketbra{y_{j_{\alpha}}}{y_{j_{\alpha}}} = \sum_{i_{\alpha} j_{\alpha}} \mu_{j_{\alpha}| i_{\alpha}} \Pcal_{j_{\alpha} i_{\alpha}}[\rho].
\end{gather}
Eq.~\eqref{eqn:stochPrep} represents the overall action of the most general classical operation an experimenter can perform. Upon measuring the outcome $x_{i_{\alpha}}$, they could reprepare the state $\ket{y_{j_{\alpha}}}$ with respective conditional probability $\mu_{j_{\alpha}|i_{\alpha}}$, \textit{i.e.}, $\mu_{j_{\alpha}| i_{\alpha}} \geqslant 0$ and $\sum_{j_{\alpha}} \mu_{j_{\alpha}| i_{\alpha}} = 1$. The influence of a sequence of such experimental manipulations on the system of interest can be expressed in terms of the joint probability distributions $p(x_{i_N};y_{j_{N-1}},x_{i_{N-1}}; \dots; y_{j_1}, x_{i_1})$. For example, the probability to measure the outcome $x_{i_N}$ at the final time step, given that the operations $P_1,\dots,P_{N-1}$ were performed at the earlier time steps, is given by 
\begin{align}
    p\left(x_{i_N}|P_{N-1},\dots,P_1\right) &= \sum_{\stackrel{i_1\dots i_{N-1}}{j_1\dots j_{N-1}}} \mu_{j_{N-1}| i_{N-1}} \cdots \mu_{j_1| i_1}
    \ p(x_{i_N}, \dots, x_{i_1} | y_{j_{N-1}},\dots,y_{j_1})\, .
    \label{eqn:JointClass}
\end{align}
Consequently, in classical physics, the state $\rho_N = \sum_{i_N} p\left(x_{i_N}|P_{N-1},\dots,P_1\right) \ketbra{x_{i_N}}{x_{i_N}}$ at $t_N$, given the intermediate experimental manipulations $P_1,\dots,P_{N-1}$, can be obtained as a convex combination of a fixed set of individual trajectories,  characterized by the joint probability distributions $p(x_{i_N};y_{j_{N-1}},x_{i_{N-1}}; \dots; y_{j_1}, x_{i_1})$.
In principle, the experimental interventions can also be temporally correlated. For example, the probability to reprepare $y_{\alpha}$ at time $t_{\alpha}$ could depend on all measurement outcomes and repreparations at times $t_{\alpha'} \leqslant t_{\alpha}$. This scenario can be accounted for by replacing the products $\mu_{j_{N-1}| i_{N-1}} \cdots \mu_{j_1| i_1}$ in Eq.~\eqref{eqn:JointClass} by a general probability distribution $\mu(y_{j_{N-1}},\dots, y_{j_1} | x_{i_{N-1}}, \dots, x_{i_1})$, that satisfies the constraints imposed by causality~\cite{pearl_causality_2009,chiribella_theoretical_2009}. Throughout most of this paper, we will only consider uncorrelated experimental interventions. We emphasize the generality of Eq.~\eqref{eqn:JointClass}; \textit{all} classical processes (with or without memory)  as well as all classical processes with interventions can be represented in this way. The existence of an underlying stochastic process is guaranteed under minor assumptions about the finite probability distributions~\cite{kolmogorov_foundations_1956,milz_kolmogorov_2017}.

We will see below that the dynamics of a quantum system with intermediate interventions can be treated in the same way, with the sole difference that the convex combinations have to be replaced by linear ones, \textit{i.e.}, coherent combinations of trajectories play a crucial role in quantum mechanics.

\subsection{Quantum stochastic trajectories} \label{sec:quantum}

In the spirit of our discussion of classical processes with interventions, we will introduce a (finite) special set of trajectories that will allow us to describe arbitrary quantum dynamics. In the classical case, we expanded $P_{\alpha}$ in terms of an extremal (convex-linear) basis $\Pcal_{j_{\alpha} i_{\alpha}}$. We will proceed analogously in the quantum case; any admissible quantum operation can be expressed as a linear combination of basis operations $\oneA_{l_{\alpha} k_{\alpha}}$, but, as we will see shortly, this linear combination does not have to be convex.

In quantum mechanics, a measurement at $t_{\alpha}$ is given by a corresponding positive operator valued measure (POVM) element $0 \leqslant \Pi_{k_{\alpha}} \leqslant \iden$, while a reprepared stated corresponds to a density matrix $\mathbf{R}_{l_{\alpha}}$. The action of a measure and reprepare operation $\oneA_{l_{\alpha} k_{\alpha}}$ on a state $\rho$ can be represented as
\begin{gather}
\label{eqn::MeasRep}
\oneA_{l_{\alpha} k_{\alpha}}[\rho] = \mathrm{Tr}\left(\Pi_{k_{\alpha}} \rho\right) \mathbf{R}_{l_{\alpha}}\, ,
\end{gather}
where the only difference to the classical case in Eq.~\eqref{eqn::MeasRepClass} is that the operators $\rho, \Pi_{k_{\alpha}}$ and $\mathbf{R}_{l_{\alpha}}$ need not commute. Consequently, there are $d^4$ instead of $d^2$ operations that span the space of of all possible manipulations of the system of dimension $d$~\cite{modi_operational_2012}. Throughout this paper, we will call an operation of this type a `causal break', because the measurement decouples the system of interest from its environment and the repreparation erases information about its causal past. Note that the last equation can be turned into a standard Kraus form by singular value decomposition of $\Pi_{k_{\alpha}}$ and $\mathbf{R}_{l_{\alpha}}$, see~\cite{milz_introduction_2017} for details.

\sloppy Just as in the classical case, an experimenter could interrogate the system of interest at times $t_N,\dots,t_1$ by implementing the measure and prepare operations $\oneA_{l_{\alpha} k_{\alpha}}$ and recording the corresponding probabilities $p(\Pi_{k_N}, \dots, \Pi_{k_1} | \mathbf{R}_{l_{N-1}}, \dots, \mathbf{R}_{l_1})$ to measure the sequence of outcomes with POVM elements $\Pi_{k_{N}},\dots,\Pi_{k_1}$, given the repreparations $\mathbf{R}_{l_{N-1}}, \dots, \mathbf{R}_{l_{1}}$. Each of these probability distribution can be considered as characterizing a particular trajectory of the system of interest; the dynamics of a system with arbitrary experimental manipulations $A_1,\dots,A_N$ can then be expressed as a linear combination of these elementary trajectories. We emphasize that this understanding of trajectories is operational rather than ontological. Trajectories are sequences of experimentally obtained measurement outcomes followed by repreparations and are not considered to describe some underlying `reality'. Unlike in the classical case, the basis set of elementary measure and reprepare trajectories is not unique. At each time $t_{\alpha}$, the experimenter could implement a different set of causal breaks; as long as they span the whole space of possible controls, respectively, the resulting joint probability distributions contain enough information to predict the dynamics of the system under arbitrary controls.

If the sets $\{ \Pi_{k_{\alpha}} \}_{k_{\alpha}=1}^{d^2}$ and $\{\mathbf{R}_{l_{\alpha}}\}_{l_{\alpha} = 1}^{d^2}$ are both informationally complete (IC), any quantum mechanically allowed operation can be represented as a completely positive (CP) map~\cite{modi_operational_2012,pollock_complete_pra}
\begin{align}
\label{eqn::ActionQuantum}
    A_{\alpha} = \sum_{k_{\alpha}, l_{\alpha} = 1}^{d^2} a_{l_{\alpha} k_{\alpha}} \oneA_{l_{\alpha} k_{\alpha}}\, , \text{where} \ a_{l_{\alpha} k_{\alpha}} \in \mathbb{R}\, .
\end{align}
On average, the influence of an experimenter is trace preserving (TP) and $A_{\alpha}$ is TP iff  $\sum_{l_{\alpha}} a_{l_{\alpha} k_{\alpha}} = 1$ in Eq.~\eqref{eqn::ActionQuantum}. Unlike in the classical case, the coefficients $a_{l_{\alpha} k_{\alpha}}$ do \textit{not} have to be positive, as long as $A_{\alpha}$ is CP; any CPTP operation $A_{\alpha}^{(\mathrm{EB})}$ that can be represented as a convex combination of causal breaks is \textit{entanglement breaking}~\cite{holevo_1998, horodecki_entanglement_2003} (though not all entanglement breaking operations can be formed as convex combinations of a single complete set of causal breaks). Classical operations are always entanglement breaking; however, in general, quantum operations preserve entanglement between the system and its environment. This implies that, by applying suitable controls, trajectories can be interfered in quantum mechanics, while they can only be statistically mixed in classical mechanics. 

We can now -- in clear analogy to the classical case -- define stochastic quantum trajectories by setting
\begin{gather}
\oneA^{(\xi)}_{N-1:1}:=(\oneA_{l_{N-1}(\xi) k_{N-1}(\xi)},\dots,\oneA_{l_{1}(\xi) k_{1}(\xi)})\, ,
\end{gather}
where $\xi\in \pk$ denotes a trajectory in the set of all $d^{4(N-1)}$ sequences of $N-1$ measurement-and-repreparation pairs, namely $\xi = \{(k_{\alpha},l_{\alpha})\}_{\alpha=1}^{N-1}$, and $(k_{\alpha}(\xi),l_{\alpha}(\xi))$ is the $\alpha$th element of the sequence $\xi$. The set $\pk$ is informationally complete in the sense that the trajectories it contains span the space of all trajectories, but, as already mentioned, it is not unique. Our notion of stochastic quantum trajectories is not directly related to the numerical tool known as quantum trajectories~\cite{daly}. We discuss other notions of trajectories in quantum mechanics in the Conclusions.

Defining the dual operators $\Delta_{k_n}$, satisfying $\Tr{\Delta_{k_n} \Pi_{l_n}} = \delta_{k_nl_n}$~\cite{modi_positivity_2012,milz_introduction_2017}, the final state $\rho_N$ at $t_N$, given that the sequence $\mathbf{A}_{N-1:1}:=(A_{N-1}, \dots,A_1)$ of CP operations was performed at times $t_{N-1},\dots,t_1$ can be obtained as a linear combination of \textit{elementary} trajectories: 
\begin{gather}
\label{eqn::QM_traj}
\rho_N(\mathbf{A}_{N-1:1}) =
\sum_{k_N} \sum_{\xi\in\pk} a^{(\xi)} p(\Pi_{k_N},\oneA^{(\xi)}_{N-1:1}) \, \Delta_{k_N}\, ,
\end{gather}
where $a^{(\xi)} := \prod_{\alpha=1}^{N-1} a_{k_{\alpha}(\xi)l_{\alpha}(\xi)}$ and $p(\Pi_{k_N},\oneA^{(\xi)}_{N-1:1}) := p(\Pi_{k_N},\dots,\Pi_{k_1(\xi)}|\mathbf{R}_{l_{N-1}},\dots,\mathbf{R}_{l_1})$ is the probability to obtain the measurement outcome corresponding to $\Pi_{k_N}$ at $t_N$ given the causal breaks $\oneA^{(\xi)}_{N-1:1}$ at the earlier time steps. As in the classical case, the sequence $\mathbf{A}_{N-1:1}$ of control operations can be temporally correlated~\cite{pollock_complete_pra, pollock_operational_prl}, both classically as well as quantum mechanically. For example, the ancilla used to implement an operation at time $t_{\alpha'}$ could be forwarded and used again to implement an operation at a later time $t_{\alpha}$. In this case, the coefficient $a^{(\xi)}$ would not be of product form, but a general function of all causal breaks that, as in the classical case, satisfies the requirements imposed by causality~\cite{chiribella_theoretical_2009}. As already mentioned, in this paper, we focus on uncorrelated control operations. 

Equation~\eqref{eqn::QM_traj} can be interpreted as follows: each possible sequence $\{\Pi_{k_N}; \mathbf{R}_{l_{N-1}}, \Pi_{k_{N-1}};\dots; \mathbf{R}_{l_1},\Pi_{k_1}\}$ of measurements and repreparations defines an elementary $N$-step trajectory of the system of interest. For an $N$-step quantum process, there are $d^{4N-2}$ elementary trajectories, while for a classical input-output process with the same number time-steps the number of trajectories would be quadratically smaller. Determining the elementary trajectories then requires an exponentially large number of quantum process tomography experiments in both quantum and classical cases. That is, recording the $d^{4N-2}$ joint probability distributions $p(\Pi_{k_N}, \dots, \Pi_{k_1}| \mathbf{R}_{l_{N-1}}, \dots,\mathbf{R}_{l_1})$ allows one to predict the final quantum state for any possible manipulations of the system that the experimenter chooses at the times $t_{N-1},\dots,t_{1}$, as a linear combination of the elementary measure and reprepare trajectories. 

Here, unlike in the standard path integral approach to quantum mechanics (or classical stochastic dynamics), we are considering active combination, through experimental manipulations, of a set of trajectories that represent the process as distinct from these manipulations. Characterising such a set is equivalent to reconstructing the process tensor for a process~\cite{pollock_complete_pra}, and is sufficient to fully describe the most general non-Markovian quantum dynamics. 

To simplify the subsequent considerations, we define \textit{elementary}, \textit{mixed}, \textit{entanglement breaking} and \textit{coherent trajectories}:
\begin{itemize}
    \item [] \textbf{Elementary trajectory} is a sequence of causal breaks.
    
    \item [] \textbf{Mixed trajectory} lies in the convex hull of a fixed informationally complete set of $N-$step elementary trajectories $\pk$.
    
    \item [] \textbf{Entanglement breaking (EB) trajectory set} is the union of all mixed trajectories corresponding to all $\pk$.
    
    \item [] \textbf{Coherent trajectory} is any trajectory that cannot be written as a convex mixture of \textit {any} set of elementary trajectories.
\end{itemize}

\begin{figure}
\centering
\includegraphics[scale=0.8]{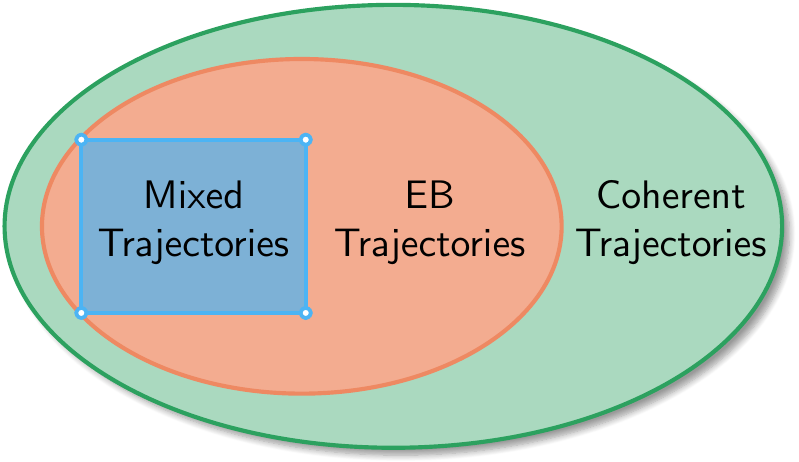}
\caption{\textit{Schematic of the different sets of trajectories.} Mixed trajectories are the ones obtained by convex combination of a fixed set of elementary trajectories (the corners of the blue rectangle). Entanglement breaking (EB) trajectories are the convex hull of \textit{all} possible elementary trajectories; they contain the set of stochastic trajectories. The coherent trajectories are those that lie outside the set of EB trajectories (the green area).}
    \label{fig::Venn}
\end{figure}

See Figure~\ref{fig::Venn} for an illustration of this definition. Classically, the set of mixed trajectories is exhaustive of all possible trajectories. For quantum processes the elementary trajectories form a linear basis for all possible events, and therefore \textit{any} trajectory can be expanded in this basis. However, this expansion will not always be convex. Each mixed trajectory consists of a sequence of of entanglement breaking operations; however, in the quantum case there are entanglement breaking trajectories that cannot be expressed as a convex combination of a \textit{fixed} set of elementary ones. Coherent trajectories are then simply trajectories that contain at least one entanglement preserving operation. Below, we show that the ability to combine elementary trajectories coherently such that the resulting trajectory lies outside the set of entanglement breaking ones allows for a greater degree of control for quantum processes. 

We have introduced `elementary quantum trajectories' in terms of outputs and inputs of causal breaks and defined the different kinds of combinations of elementary trajectories that quantum mechanics allows for. However, we have not yet discussed the underlying dynamics between an input, say $\mathbf{R}_{l_{\alpha}}$ and the output probability corresponding to a subsequent measurement $\Pi_{k_{\alpha+1}}$, \textit{i.e.},  $p(\Pi_{k_{\alpha+1}} | \mathbf{R}_{l_{\alpha}})$. We will show in the next section that this dynamics -- and consequently any elementary trajectory -- is governed by a set of CPTP maps $\Phi_{\alpha+1:\alpha|\mathbf{A}_{\alpha:1}}$ that are conditioned on the past inputs and outputs $\mathbf{A}_{\alpha:1}$. To do so, we will look at the dilation of our stochastic process, \textit{i.e.}, the unitary dynamics of the system and its environment.

\subsection{Dilation of quantum stochastic dynamics}

\begin{figure}
    \centering
    \includegraphics{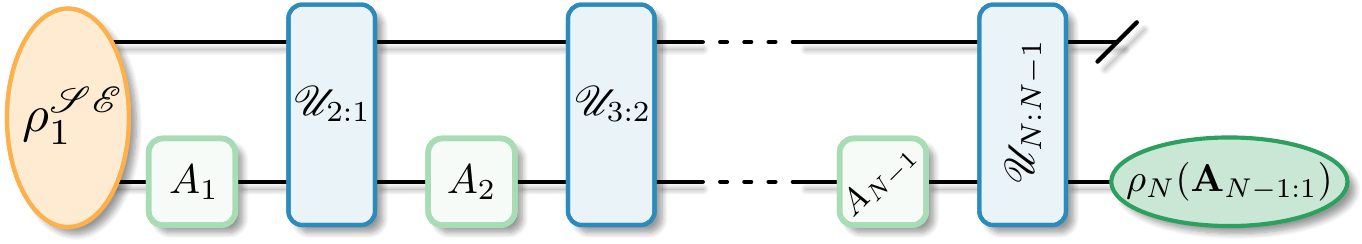}
    \caption{\textit{Dilated circuit diagram of a quantum process.} The final $\Sys \Env$ state $\rho_N^{\Sys \Env}$ is is the result of a cocatenation of unitaries and control operations acting on the initial state $\rho_1^{\Sys\Env}$ (see Eq.~\eqref{eq:iterating}). The final state of the system is given by $\rho_N(\mathbf{A}_{N-1:1}) = \Tr{\rho_N^{\Sys \Env}}$. }
    \label{fig::PolymerCircuit}
\end{figure}

In addition to the trajectory description, for open quantum dynamics one can also describe the evolution of the system of interest as emerging from a joint unitary system-environment dynamics~\cite{chiribella_theoretical_2009,pollock_complete_pra}. If we denote the initial state of the composite system-environment state as $\rho^{\Sys\Env}_1$, the final state after $N-1$ local (\textit{i.e.}, only acting on the system) operations $A_{\alpha}$ is given by
	\begin{gather}
	\label{eq:iterating}
		\rho^{\Sys\Env}_N = \left(\prod_{\alpha=1}^{N-1}\Ucal_{\alpha+1:\alpha} \circ\left(A_{\alpha} \otimes\I_{\Env}\right)\right)\left[\rho^{\Sys\Env}_1\right]\, ,
	\end{gather}
where $\Ucal_{\alpha+1:\alpha}[\omega^{\Sys\Env}_{\alpha}] = U_{\alpha+1;\alpha} \omega^{\Sys\Env}_{\alpha} U^{\dagger}_{\alpha+1;\alpha} := \omega^{\Sys\Env}_{\alpha+1}$ is the system-environment unitary evolution between $t_{\alpha}$ and $t_{\alpha+1}$, and $\I_{\Env}$ is the identity map on the environment (see Fig.~\ref{fig::PolymerCircuit}). While less commonly considered than in the quantum case, classical stochastic processes also have a `dilated' representation, similar to that in Eq.~\eqref{eq:iterating}. In this case, the $\Sys\Env$ unitary becomes a simple permutation. 
Eq.~\eqref{eq:iterating} is a dynamical representation of the trajectory picture, in the sense that the final system state $\rho_N = \partTr{\Env}{\rho^{\Sys\Env}_N}$ coincides with the state $\rho_N(\mathbf{A}_{N-1:1})$ in Eq.~\eqref{eqn::QM_traj}.  A causal break leaves the system and the environment in a product state, \textit{i.e.}, the local action of $\oneA_{l_{\alpha} k_{\alpha}}$ yields
\begin{gather}
    \left(\oneA_{l_{\alpha} k_{\alpha}} \otimes \Ical_\Env\right)[\omega^{\Sys\Env}_{\alpha}] = \mathbf{R}_{l_{\alpha}} \otimes \partTr{S}{(\Pi_{k_{\alpha}}\otimes \iden_\Env)\omega^{\Sys\Env}_{\alpha}}\, .
\end{gather}

Consequently, when all operations in Eq.~\eqref{eq:iterating} are causal breaks, the dynamics of the system between any two time steps $t_{\alpha}$ and $t_{\alpha+1}$ is given by a CP map $\Phi_{\alpha+1:\alpha| \oneA^{(\xi)}_{\alpha:1}}$ that depends on all causal breaks (the trajectory) up to times $t_{\alpha'} \leqslant t_{\alpha}$ and the measurement outcome at $t_{\alpha}$, as well as the initial state $\rho^{\Sys\Env}_1$. That is, the dynamics from $t_{\alpha}$ to $t_{\alpha+1}$ is \textit{trajectory dependent}. Hence, we can write the overall dynamics of the system as a concatenation of CP maps acting on the initial state of the system:
\begin{align}
\label{eqn::CausalBreakDyn}
\rho_N(\oneA_{N-1:1}^{(\xi)})	=&	\partTr{\Env} {\prod_{\alpha=1}^{N-1}\Ucal_{\alpha+1:\alpha}\circ\left(\oneA_{l_{\alpha}(\xi) k_{\alpha}(\xi)} \otimes \I_{\Env}\right) \left[ \rho^{\Sys\Env}_1 \right]} \\ \nonumber
=&	\left(\prod_{\alpha=1}^{N-1} \Phi_{\alpha+1:\alpha|\oneA^{(\xi)}_{\alpha:1}}\circ \oneA_{l_{\alpha}(\xi) k_{\alpha}(\xi)}\right)\left[\rho_1\right],
\end{align}
where $\rho_1 = \partTr{\Env} {\rho^{\Sys\Env}_1}$ and $\rho_N(\oneA_{N-1:1}^{(\xi)})$ is a subnormalised quantum state. The dynamics of the system is \textit{Markovian} (or \textit{memoryless}) iff the maps $\Phi_{\alpha+1:\alpha | \oneA^{(\xi)}_{\alpha:1}}$ do not depend on the history (trajectory), \textit{i.e.}, $\Phi_{\alpha+1:\alpha | \oneA^{(\xi)}_{\alpha:1}} = \Phi_{\alpha+1:\alpha} \ \forall \ \xi, \alpha$~\cite{pollock_operational_prl}. This requirement is strictly stronger than CP divisibility~\cite{PhysRevLett.105.050403, pollock_operational_prl}.

If all implemented operations $\{A_{\alpha}\}_{\alpha=1}^{N-1}$ are entanglement breaking, the corresponding final state can be written as the result of a convex combination of dynamics of the form~\eqref{eqn::CausalBreakDyn} for some sets of IC POVMs $\{\Pi_{k_{\alpha}}\}_{k_{\alpha} = 1}^{d^2}$ and IC states $\{\mathbf{R}_{l_{\alpha}}\}_{l_{\alpha}=1}^{d^2}$. Combining Eqs.~\eqref{eqn::QM_traj} and~\eqref{eqn::CausalBreakDyn}, a \textit{general} CPTP operation $A_{\alpha}$ can be written as a linear (but not convex) combination of causal breaks, and one obtains the corresponding final state
\begin{align}
\rho_N(\mathbf{A}_{N-1:1})	\label{eqn::TrajSum}
&=	\sum_{\xi\in\pk}a^{(\xi)}\left(\prod_{\alpha=1}^{N-1}\Phi_{\alpha+1:\alpha|\oneA^{(\xi)}_{\alpha:1}}\circ\oneA_{l_{\alpha}(\xi)k_{\alpha}(\xi)}\right)\left[\rho_1 \right].
\end{align}

This implies that the final quantum state for any sequence of intermediate manipulation can be calculated, once a sufficient number of multi-time correlation functions has been determined. In a more succinct way, this fact can be expressed in terms of a~\textit{process tensor}~\cite{modi_operational_2012, pollock_complete_pra, milz_introduction_2017}, an operationally well-defined \textit{quantum comb}~\cite{chiribella_transforming_2008, chiribella_quantum_2008, chiribella_theoretical_2009} that maps sequences of CP maps to final quantum states. We will employ this description below to investigate three-step processes, \textit{i.e.}, processes where the system is manipulated/interrogated at three points in time.

\section{Interfering stochastic quantum trajectories} \label{sec:interfering}

In the notation of Eq.~\eqref{eqn::TrajSum}, the dynamics of the system becomes a summation over elementary measure and reprepare quantum trajectories. Indeed, this represents a discrete time path summation analogous to the continuous stochastic-path integration employed in Refs.~\cite{Chantasri2013, Chantasri2015}. We reiterate that the dynamics is non-Markovian iff any of the individual trajectories exhibits a history dependence. If the experimenter can perform operations that preserve entanglement between the system and its environment, the summation in Eq.~\eqref{eqn::TrajSum} is necessarily non-convex. In what follows, we will use both non-Markovianity and coherence as resources. Here, it is important to distinguish between two types of coherence; one is the coherence in the stochastic process itself, \textit{i.e.}, its capability to preserve coherence in the state of the system of interest. For example, a classical stochastic process would destroy coherence exhibited by the system state in some basis, and could not be used for quantum mechanical decoupling scenarios (see below). The other type of coherence is that at the disposal of the experimenter as soon as they can perform operations that preserve entanglement. In this case, the overall dynamics lies outside the set of convex combinations of elementary trajectories, allowing for their coherent interference.

To make these statements concrete we now construct an explicit example where history dependent CPTP maps emerge, and show how stochastic quantum trajectories can be usefully interfered. To avoid unnecessary complexity we confine ourselves to a three time-step process, \textit{i.e.}, a process where the system of interest is prepared at time $t_1$, manipulated at time $t_2$ and measured at time $t_3$. We consider a game with three players involved: Alice, Bob, and Charlie. At time $t_1$ Alice prepares a quantum state and sends it to Charlie via Bob. By the time the quantum system reaches Bob at $t_2$, its state has changed. Bob may not know which state was prepared by Alice, but he is allowed to perform an operation on the quantum system with the goal that Charlie receives a state at $t_3$ which has high fidelity with the state prepared by Alice. With this example, we will show that when the dynamics are non-Markovian, coherent quantum trajectories can cancel decoherence effects.

\subsection{Three-step process} 
\label{subsec::Three_step}

We begin by constructing the elementary trajectories for our game. Let the total initial state be of product form: $\rho^{\Sys\Env}_1 = \rho_1 \otimes \rho^\Env_1$, where the experimenter can freely prepare the system state $\rho_1$\footnote{The initial product form is chosen for convenience, but is not necessary for the arguments of the example to hold.}. We can expand Alice's initial state in terms of a basis as $\rho_1 = \sum_{l_1} a_{l_1} \mathbf{R}_{l_1}$. For each $\mathbf{R}_{l_1}$, after an interaction with the environment (given by the $\Sys\Env$ unitary $U_{2:1}$), the system is in state $\rho_{2|l_1}$. Next, Bob performs a causal break operation $\oneA_{l_2 k_2}$ on the system, and it subsequently interacts with the environment again to yield $\rho_{3|l_2k_2l_1}$. The dynamics from $t_1$ to $t_2$ is governed by a usual CPTP map $\Phi_{2:1}$~\cite{SudarshanMatthewsRau61, kraus_general_1971}, which depends on $\rho_1^{\Env}$ but not on the choice of $\rho_1$. The dynamics from $t_2$ to $t_3$ is governed by a set of conditional maps that depend on the choice of initial state $\rho_1$ and the measurement outcome $k_2$ of the causal break.

\begin{figure}
    \centering
    \includegraphics[scale=0.7]{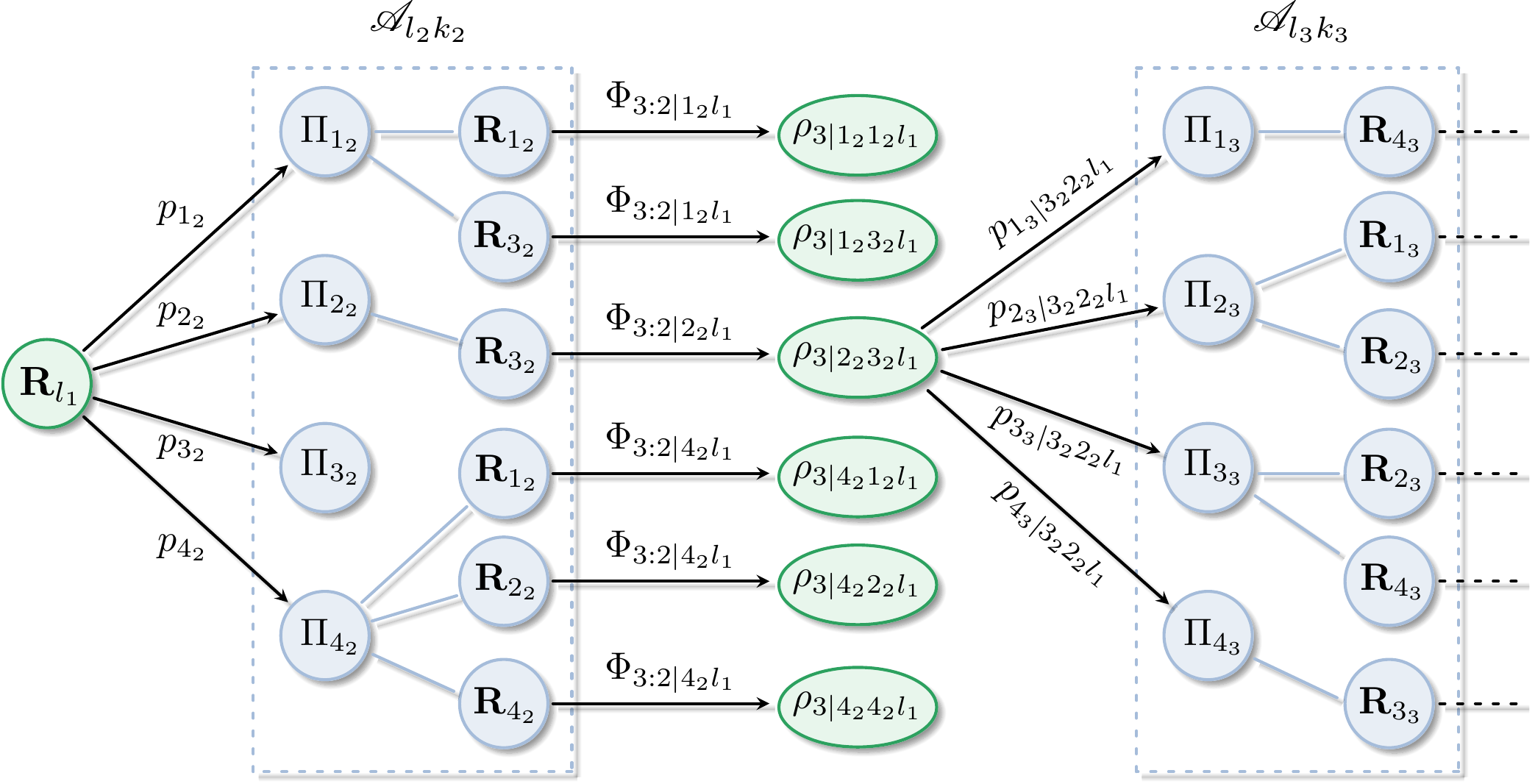}
    \caption{\textit{Elementary quantum trajectories.} For a fixed initial system state, each sequence of causal breaks leads to a different trajectory. For a qubit, there are $16$ causal breaks that span the space of control operations. The initial measurement outcome occurs with probability $p_{k_2}$, the second measurement outcomes with probability $p_{3|l_2k_2\rho_1}$, etc.. The map $\Phi_{3:2|k_2 \rho_1}$ depends on the respective trajectory it belongs to. For compactness, only some of the possible elementary trajectories are depicted.} 
    \label{fig:my_label}
\end{figure}

Explicitly, these CPTP maps can be written as
\begin{align}
& \Phi_{2:1}[\mathbf{R}_{l_1}] = \partTr{\Env}{U_{2:1} \mathbf{R}_{l_1}\otimes \rho^{\Env}_1 U^\dag_{2:1}} =\rho_{2|l_1}
\quad \mbox{and} \\
& \Phi_{3:2|k_2l_1}[\mathbf{R}_{l_2}] = \partTr{\Env}{U_{3:2} \mathbf{R}_{l_2} \otimes \rho^\Env_{2|k_2l_1} U^\dag_{3:2}}
=\rho_{3|l_2k_2l_1},
\end{align}
where $\rho^\Env_{2|k_2l_1} = \partTr{\Sys}{\Pi_{k_2} \rho^{\Sys\Env}_{2|l_1}}/p_{k_2|l_1}$ and $p_{k_2|l_1} = \Tr{\Pi_{k_2} \rho^{\Sys\Env}_{2|l_1}}$, with $\rho^{\Sys\Env}_{2|l_1}$ being the correlated state at $t_2$. Subsequently, Charlie receives the state $\rho_{3|l_2k_2l_1}$ at $t_3$, which is a function of the basis state $\mathbf{R}_{l_1}$ at $t_1$ and the causal break at $t_2$. With knowledge of the trajectory dependent maps $\Phi_{3:2|k_2l_1}$, the final system state for any initial state $\rho_1$ (prepared by Alice) and intermediate manipulation $\Acal_2$ (performed by Bob) can be calculated by expanding $\rho_1$ and  $\oneA_2$ in a suitable basis of density operators and causal breaks, respectively, and using Eq.~\eqref{eqn::TrajSum}; a break down in terms of a particular trajectory is depicted in Figure~\ref{fig:my_label}.

\begin{figure}
\centering
\includegraphics[scale=0.7]{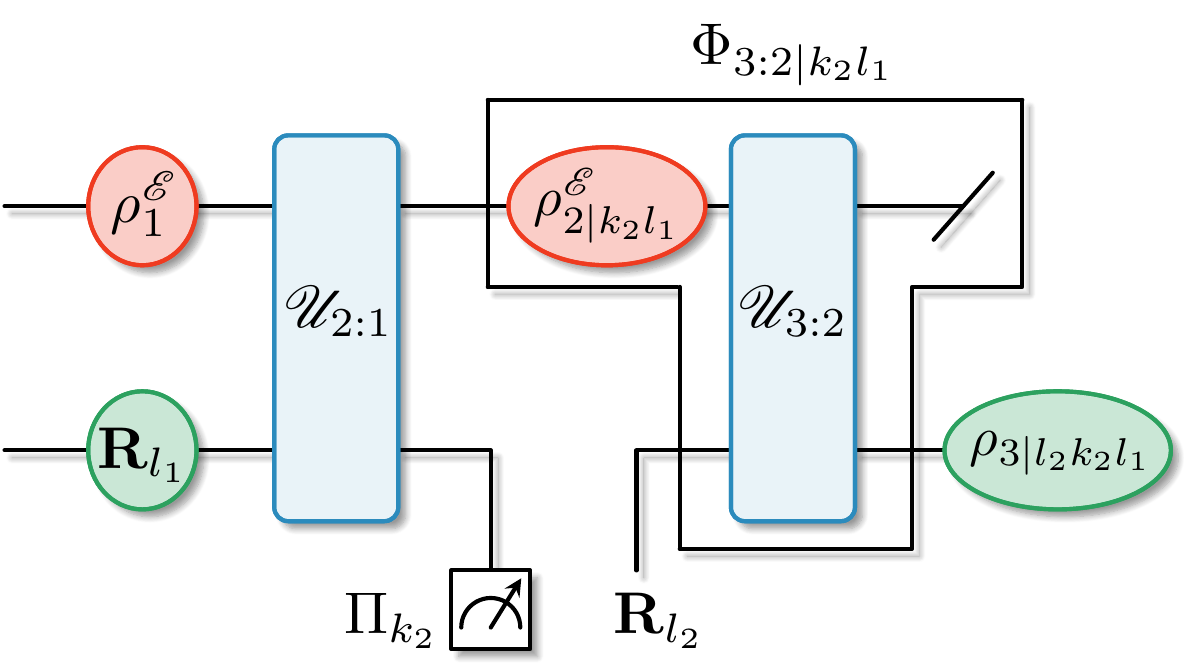}
\caption{\textit{Three-step process with a causal break}. The causal break at $t_2$ leaves $\Sys$ and $\Env$ in a product state, where the state $\rho_{2|k_2l_1}^{\Env}$, in general, depends on the measurement outcome and the initial system state. The subsequent dynamics can be described on the level of the system by a CP map $\Phi_{3:2|k_2\rho_1}$ that is \textit{trajectory dependent}; it depends on the initially prepared system state as well as the causal break that was performed. The final system state is given by $\rho_{3|l_{\alpha} k_{\alpha} \rho_1} = \Phi_{3:2|k_2\rho_1}[\mathbf{R}_{l_{\alpha}}]$.}
\label{fig::three_step}
\end{figure}

We further simplify the playing field by looking at a restricted class of total Hamiltonians generating the $\Sys\Env$ dynamics of the form $\opH = \opS\otimes\opB$, and we take $\Sys$ to be a qubit with Hamiltonian $\opS=\sigma_z/2$. Any initial state that does not commute with $\sigma_z$ will become more mixed under the subsequent dynamics. For instance, let Alice prepare $\rho_1 = \ket{\psi_1} \bra{\psi_1}$, with $\ket{\psi_1} = \mu_1 \ket{0}+\nu_1 \ket{1}$. Then the state that Bob will receive has the form
\begin{gather}\label{eq:stateat2}
\rho_2 = \Phi_{2:1}\left[
\left(\begin{matrix} 
|\mu_1|^2 & \mu_1\nu_1^* \\ \mu_1^* \nu_1 & |\nu_1|^2\\
\end{matrix}\right)\right]
=\left(\begin{matrix}
|\mu_1|^2 & \mu_1\nu_1^* f_{2:1}^* \\ \mu_1^* \nu_1 f_{2:1}& |\nu_1|^2\\
\end{matrix}\right)
\quad \mbox{where} \quad
f_{2:1} := \mathrm{Tr}\left(\rho_1^\Env e^{-i \opB(t_2-t_1)}\right).
\end{gather}
Expanding the $\Env$ Hamiltonian in its eigenbasis as $\opB=\sum_\gamma b_\gamma \ket{\gamma}\!\bra{\gamma}$, we get $f_{2:1} = \sum_\gamma \braket{\gamma | \rho^\Env_1 | \gamma} e^{-i b_\gamma (t_2-t_1)}$. Therefore, unless $\rho^\Env_1 = \ket{\gamma}\!\bra{\gamma}$ or there are degeneracies in $\opB$, we have $|f_{2:1}| < 1$.

The fidelity of $\rho_2$ with respect to the initial state $\rho_1$ is $F(\rho_1,\rho_2) = \braket{\psi_1 |\rho_2|\psi_1} =1 - 2|\mu_1|^2 |\nu_1|^2 [1 - \text{Re}(f_{2:1})]$. Here we have used the fact that $\rho_1$ is pure. If $\text{Re}(f_{2:1}) < 1$, it means that the state transfer was not perfect. However, if $\text{Re}(f_{2:1}) <1$ but $|f_{2:1}| = 1$ then the original state can be recovered by action of a local unitary on the system alone.

Now, suppose Bob applies a causal break for his operation and sends a state $\ket{\psi_2}$ to Charlie. The fidelity between the state Charlie receives and the state Bob sent can be computed in the same way as above, since the dynamics from Bob to Charlie is governed by the same $\Sys\Env$ Hamiltonian as that from Alice to Bob. Therefore, the best case scenario is that Bob knows the state Alice has prepared; he then simply reprepares this state and sends it Charlie. However, without prior information, he cannot know the state Alice is preparing on average from a single causal break. As a result, $F(\rho_1,\rho_3) \leqslant F(\rho_1,\rho_2)$ for any causal break Bob performs. By convexity of fidelity, the function $F(\rho_1,\rho_3)$ will always decrease for any operation Bob makes that is a convex mixture of causal breaks, \textit{i.e.}, entanglement breaking channels. This does not hold if Bob is allowed to interfere elementary trajectories in a non-convex way.

Concretely, we can express any CPTP map $\oneA_2$ performed by Bob as a linear sum of the causal breaks above: $\oneA_{2} = \sum_{l_2 k_2}  a_{l_2 k_2} \oneA_{l_2k_2} $. The corresponding state at $t_3$ is $\rho_{3|\oneA_2} = \sum_{l_2k_2l_1} a_{l_2k_2}a_{l_1} \rho_{3|l_2k_2l_1}$, where the coefficients $a_{l_1}$ correspond a decomposition of Alice's state as above. When $a_{l_2k_2}>0$, the fidelity between the states at $t_1$ and $t_2$ is less than one. However, not all CPTP maps can be expressed with positive $a_{l_2k_2}$. For instance any extremal map, including unitary maps such as the identity map, where Bob simply leaves the state he receives unchanged before sending it out again, cannot be expressed as a convex mixture of causal breaks. That is, to express such operations, we need to use nonpositive coefficients $a_{l_2k_2}$, resulting in interference between elementary trajectories. 

A more interesting choice for the CPTP map at $t_2$ is the \textsc{not} gate: $\textsc{not}\rho = \sigma_x\rho\sigma_x$. This too is a unitary operation and leads to interference. Using 
\begin{gather}
\sigma_x \otimes \iden e^{-i \sigma_z/2 \otimes\opB t}\sigma_x \otimes \iden = e^{i \sigma_z/2 \otimes\opB t} \quad \mbox{and} \quad \sigma_x \sigma_x = \iden
\end{gather}
we find that $\rho_{3|\oneA_2=\textsc{not},\rho_1} = \sigma_x \rho_1 \sigma_x$. That is, we have complete constructive interference and the final state $\rho_3$ is the \textsc{not} of the initial state $\rho_1$, thus applying a $\sigma_x$ operation at $t_3$ makes $\rho_3 = \rho_1$. The model above could fine tuned so that the loss of information is uniform in time, see Ref.~\cite{PhysRevA.92.022102} for an example. Such a phenomenon is not allowed in classical stochastic processes. To observe it, we need both kinds of coherence mentioned at the start of this section: non-Markovianity and coherent operations that cannot be expressed as convex mixture of elementary trajectories. Of course, we have not yet quantified the former, and we will now do so.

\subsection{Measuring non-Markovianity} 

\begin{figure}
    \centering
    \includegraphics[scale=1.1]{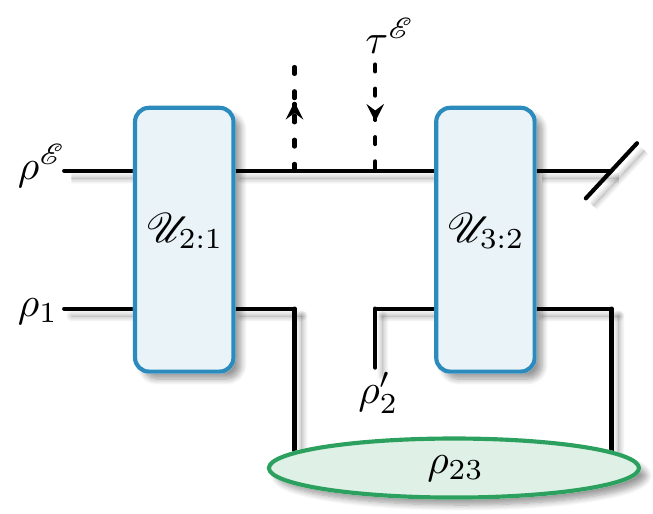}
    \caption{\textit{Mutual Information.} Bob stores the state that he receives from Alice and feeds a fresh state $\rho_2'$ forward. If there are memory effects, the mutual information between Bob and Charlie can be non-vanishing. The memoryless case is depicted by dotted lines. In this case, the state of the environment is reset to $\tau^\Env$ after the first time step, and Bob and Charlie are entirely uncorrelated.}
    \label{fig::Mutual_Info}
\end{figure}

The non-Markovianity itself manifests in the temporal correlations between Alice, Bob, and Charlie. To see this let us consider a slight variation on the setup above. As before Alice prepares a state $\rho_1$ and correspondingly we call Bob's received state $\rho_2$. Bob then saves this state locally and sends an independently chosen state $\rho'_2$ to Charlie (see Fig.~\ref{fig::Mutual_Info}). As before, let's call the received state of Charlie $\rho_3$. We want to quantify the correlations -- and hence the non-Markovianity of the process -- between the two states that Bob and Charlie possess at the end of the protocol, \textit{i.e.}, $\rho_2$ and $\rho_3$\footnote{For simplicity of notation, we will omit the explicit trajectory dependence of the states for the most part of this section.}. That is, the quantum mutual information $I(B:C)_{\rho_{23}} = S(\rho_2)+S(\rho_3)-S(\rho_{23})$, where $S(\rho)=-\mathrm{Tr}[\rho \log(\rho)]$ is the von Neumann entropy.

First, let us consider the Markovian, \textit{i.e.}, memoryless, version of the above set up. In this case the mutual information $I(B:C)_{\rho_{23}}$ is vanishing, since the dynamics connecting Bob and Charlie is independent of the choice of Alice's initial state and Bob's causal break. As a consequence, Bob cannot do anything so that $F(\rho_3,\rho_1) > F(\rho_2,\rho_1)$ (assuming he doesn't know which initial state Alice prepared). On the other hand, when the dynamics are non-Markovian, Bob will have correlations with Charlie. He can use these correlations to help out Alice and Charlie in their communication task. 

Even in the non-Markovian case, where $I(B:C)_{\rho_{23}}>0$, it is crucial that Bob is able to generate coherent trajectories, \textit{i.e.}, trajectories that are not convex mixtures of causal breaks\footnote{It is easy to show that Bob should only apply extremal maps, which include unitary operations and some entanglement breaking maps. In fact, there are examples where Bob is better off making an entanglement breaking operation. However, this is not the case for our choice of interaction Hamiltonian.}. To see this, let us compute the correlations between Bob and Charlie when Bob chooses to implement an operation that is a convex mixture of causal break operations. That is, Bob makes a measurement on $\rho_2$ and observes outcome $k_2$ with probability $p_{k_2}$. Conditioned on this outcome he sends Charlie $\mathbf{R}_{l_2}$ with probability $p_{l_2|k_2}$, who subsequently receives $\rho_{3|l_2 k_2}$. At the end of this protocol, Bob does not hold a quantum system. He only has a record of his measurement outcomes and subsequent preparations, and consequently, his state is given by $\rho_2 = \sum_{l_2 k_2} p_{l_2|k_2} p_{k_2} \ketbra{l_2 k_2}{l_2 k_2}$, where $\{\ket{l_2 k_2}\}$ are orthogonal `flag' states of a classical register; the combined state of Bob and Charlie is of the form $\rho_{23} = \sum_{l_2 k_2} p_{l_2|k_2} p_{k_2} \ketbra{l_2 k_2}{l_2 k_2} \otimes \rho_{3|l_2 k_2}$. The correlations between Bob and Charlie in this case are given by
\begin{gather}
I_{\rm cl}(B:C)_{\rho_{23}} =  S \left( \textstyle{\sum_{l_2 k_2 }}
p_{k_2} p_{l_2|k_2} \rho_{3|l_2 k_2} \right) - \textstyle{\sum_{p_{l_2 k_2 }}}
p_{k_2} p_{l_2|k_2} S \left( \rho_{3|l_2k_2} \right).
\end{gather}
This correlation resembles the classical correlation defined in Ref.~\cite{henderson}. That is, Bob and Charlie only share classical correlations in this case.

Finally, returning to the case where Bob is able to save the quantum state $\rho_2$ in his lab and send Charlie $\rho'_2$, \textit{i.e.}, he is able to perform operations that are non-convex combinations of causal breaks. The corresponding mutual information between the two will be a quantum mutual information $I_{\rm q}(B:C)_{\rho_{23}}$. Using, concavity of entropy (and convexity of mutual information) it is easy to show that $I_{\rm q} \geqslant I_{\rm cl}$~\cite{rmp}. In fact, the difference between these two mutual informations is exactly quantum discord. In short, we have quantitatively shown there are more correlations (information) in a non-Markovian quantum process than its classical counterpart. Moreover, in the previous section we showed that these correlations can be a resource. If used correctly, we can retrieve the information stolen by the environment.

\section{Coherent control for \texorpdfstring{$\opH=\opS\otimes\opB$}{}}\label{sec:simple}

In this section, we will examine in more detail how a coherent combination of trajectories can reduce, or even eliminate, decoherence. In particular, we will supply the considerations of the previous section with a geometrical interpretation that also allows for more quantitative statements. To this end, as in the previous section, we restrict ourselves to the case of $\Sys\Env$ Hamiltonians of the form   $\opH=\opS\otimes\opB$. We emphasize that the resulting reduced dynamics of the system of interest for this type of Hamiltonian is non-Markovian; indeed, any nontrivial time-independent system-environment Hamiltonian will always lead to non-Markovian dynamics~\cite{pollock_operational_prl, PhysRevA.92.022102}. We already know from the previous section that the possibility of coherent, entanglement preserving trajectories can be a resource for the control of a system of interest. Here, we show with an example how this resource, in conjunction with memory effects of the underlying dynamics, is employed explicitly in the suppression of decoherence effects, \textit{i.e.}, in the field of dynamical decoupling~\cite{Viola1999}.

It is straightforward to show that, for our choice of Hamiltonian, the reduced dynamics of the system is unital, \textit{i.e.}, all conditioned dynamical maps leave the maximally mixed state invariant. We show this in detail in \ref{app::unital}. Unital maps have a simple geometrical interpretation; by representing states of a $d$-dimensional system in terms of generalized Pauli-matrices, the state space of density matrices can be mapped to a convex subset of $\mathbb{R}^{d^2-1}$~\cite{bengtsson_geometry_2007}. The action of a unital CPTP map can then be decomposed into a rotation and a scaling of the state space. We detail the mathematics of this simple picture in \ref{appen::Decom}. For example, a unital map acting on a qubit will simply rotate and shrink the Bloch sphere ~\cite{bengtsson_geometry_2007, KingRuskai2001,fujiwara_one--one_1999}. 

In general, the rotational part corresponds to unitary dynamics, while the scaling part contains all information about decoherence. It can therefore -- just like the decrease of fidelity in the previous section -- be considered as a measure of irreversibility~\cite{FattahPreprint, KingRuskai2001}. In this geometrical picture, the task of maximizing the fidelity between the input and output state via intermediate manipulations now translates to minimizing the scaling part of the resulting map; a vanishing scaling part implies that the overall dynamics is unitary, and hence reversible. In the more general, non-unital case, a similar decomposition can be performed~\cite{FattahPreprint}, however the interpretation in terms of irreversibility is not clear, due to the emergence of an additional translational part. We emphasize, though, that even for the non-unital case a vanishing scaling part would imply unitary dynamics. 

Let $\opS=\sum_{\mu}^d s_\mu\opP_\mu$ and $\opB=\sum_{\gamma}^{d_\Env} b_{\alpha}\opP^{\Env}_\gamma$ be the spectral decompositions of operators $\opS$ and $\opB$, where $\opP_\mu$ and $\opP^{\Env}_\gamma$ denote the corresponding spectral projection operators, and $d$ and $d_\Env$ are the dimensions of the corresponding Hilbert spaces. 
Subsequent to a causal break $\oneA_{l_{\alpha} k_{\alpha}}$  at the time step $t_{\alpha}$,
 the reduced evolution can be rewritten as
\begin{gather}
\Phi_{\alpha+1:\alpha|k_{\alpha},\mathbf{A}_{\alpha-1:1}}\left[\mathbf{R}_{l_{\alpha}}\right] = \sum_\gamma p_{k_{\alpha},\gamma} e^{-it b _\gamma \opS}{\mathbf{R}_{l_{\alpha}}}e^{it b _\gamma \opS}
= \sum_{\mu\mu'}\sum_\gamma p_{k_{\alpha},\gamma} e^{-it b _\gamma\omega_{\mu\mu'} }\proj{\mu\mu'}{\mathbf{R}_{l_{\alpha}}},\label{eq:SubspaceEvolution}
\end{gather}
where $\proj{\mu\mu'}{\cdot}=\opP_\mu\left(\cdot \right)\opP_{\mu'},$ $\omega_{\mu\mu'}=\left(s_\mu-s_{\mu'} \right),$ and $p_{k_{\alpha},\gamma} = \Tr{\opP^{\Env}_\gamma\rho_{\alpha|k_{\alpha},\mathbf{A}_{\alpha-1:1}}^{\Env}}$. For compactness, we omit the dependence of $p_{k_{\alpha},\gamma}$ on the history $\mathbf{A}_{\alpha-1:1}$. The final state $\rho_{\alpha|k_{\alpha},\mathbf{A}_{\alpha-1:1}}^{\Env}$ is history dependent, `containing' the influence of the past controls $\mathbf{A}_{\alpha-1:1}$ on the future dynamics. From Eq.~\eqref{eq:SubspaceEvolution}, we can see that the history dependence of the conditioned maps $\Phi_{\alpha+1:\alpha|k_{\alpha},\mathbf{A}_{\alpha-1:1}}$ is entirely encoded in the coefficients $p_{k_{\alpha},\gamma}$.

In the previous section, we showed that for an $\Sys\Env$ Hamiltonian of the form $\opH=\opS\otimes\opB$, a \textsc{not} gate can be used to recover the original system state despite the decohering influence of the environment when the system is a qubit. In the general, $d$-dimensional case, perfect recovery can also be achieved; instead of a single application of the \textsc{not} gate, the shift operator $\op{G}$ and its powers have to be applied at $d$ equally spaced time steps. In the eigenbasis of $\op{S}$, the shift operator can be written as the $d\times d$ matrix
	\begin{gather}
	 	\op{G} = \left(\begin{array}{ccccc}
	 	0 & 1 & 0 & \ldots & 0 \\ 
	 	0 & 0 & 1 & \ldots & 0 \\ 
	 	\vdots & \vdots & \vdots & \ddots & \vdots \\ 
	 	0 & 0 & 0 & \ldots & 1 \\ 
	 	1 & 0 & 0 & \ldots & 0
	 	\end{array}  \right)\, .
	 \end{gather} 
Perfect decoupling can be achieved by implementing the operators $\op{G},\op{G}^2,\dots,\op{G}^d$ at equidistant points in time. 
In detail, we have \footnote{Here, the order of the unitary dynamics and the control operations slightly differs from the order in Sec.~\ref{subsec::Three_step}. This choice for convenience does not impact the arguments we make.}
	\begin{gather}
	\label{eqn::PermDecoup}
		\Phi\left[ \rho \right] := \partTr{\Env}{\left(\mathcal{G}^d\otimes\I_{\Env}\right)\circ\ldots\circ\left(\mathcal{G}^2\otimes\I_{\Env}\right)\circ\Ucal\circ\left(\mathcal{G}\otimes\I_{\Env}\right)\circ\Ucal\left[\rho\otimes\rho_1^{\Env} \right]} = \mathcal{V}[\rho] ,
	\end{gather}
where $\mathcal{G}[\cdot] \equiv \op{G}(\cdot)\op{G}^\dagger$ and $\Ucal[\cdot] = e^{-i t \opS\otimes \opB} (\cdot) e^{i t \opS\otimes \opB},$ and $\mathcal{V}$ is a fixed unitary map. As the state $\Phi\left[ \rho \right]$ will be unitarily equivalent to the initial state $\rho$, there is no overall information transfer between the system and the environment, \textit{i.e.}, at the final time the system is perfectly decoupled from its surroundings. For instance, in the three-step qubit process in the previous section, the operator $\op{G}$ is selected to be $\sigma_x$ where $\op{G}^2 =\iden$. While Eq.~\eqref{eqn::PermDecoup} can be proven in a straightforward fashion \cite{FrancescoAugusto2005}, in what follows, we will show, how to obtain this result by interfering elementary trajectories, such that the scaling part of the resulting map vanishes.

To do so, in a first step, we derive the scaling part of the system dynamics for a single step process, \textit{i.e.}, for the conditional maps defined in Eq.~\eqref{eq:SubspaceEvolution}. 
Following the procedure detailed in \ref{appen::Decom}, we employ the notion of scaling parameters $l^{(k_\alpha)}_{\mu \mu'}(t)$; intuitively, $e^{-l^{(k_\alpha)}_{\mu \mu'}(t)}$ can be thought of as the shrinking factor of a particular basis element $\vert\mu\rangle\langle\mu'\vert$. For the conditional map $\Phi_{\alpha+1:\alpha|k_{\alpha},\mathbf{A}_{\alpha-1:1}}$ defined in Eq.~\eqref{eq:SubspaceEvolution}, one obtains scaling parameters of the form 
\begin{gather}
\label{eqn::scaling_cond}
\ell^{(k_{\alpha})}_{\mu\mu'}(t)=-\dfrac{1}{2}\ln\displaystyle\left\vert  \sum_{\gamma,\gamma'}   p_{k_{\alpha},\gamma} p_{k_{\alpha},\gamma'}\cos\left( \Omega_{\gamma\gamma'}\omega_{\mu\mu'}t  \right)\right\vert,
\end{gather}
where $\Omega_{\gamma\gamma'}=b_\gamma-b_{\gamma'}.$ 
The explicit form of the scaling parameter depends on the selected basis; for convenience and we use the eigenbasis of $\op{S}$. 
The relation of the scaling parameters to decoherence can be made manifest. For example, for a unital map $\Psi$, the linear entropy $S_L(\eta) := 1-\tr{\eta^2}$ of a state $\eta = \Psi[\rho]$ can be simply expressed as $S_L(\eta) = \dfrac{d-1}{d}-\sum_{k=1}^m e^{-2\ell_k(t)}\vert\vec{x}(\rho)\vert^2$ where $k=1,\ldots,m$ denote diagonal subspaces (\textit{i.e.}, they refer to parameters of the type $l_{\mu \mu}^{(k_\alpha)}$ in Eq.~\eqref{eqn::scaling_cond}), $\ell_k(t)$ is the corresponding scaling parameter for $\Psi$, and $\vec{x}(\rho)$ is the generalised Bloch vector of the state $\rho$~\cite{FattahPreprint}.

For an elementary trajectory, the scaling parameter $\ell^{(k_{\alpha})}_{\mu\mu'}$ expresses the information transfer between system and environment for a single step of the experiment. This parameter depends on the energy spacing $\omega_{\mu\mu'}$ (which cannot be controlled by the experimenter), as well as the conditional state of the environment at $t_{\alpha}$ which is reflected in the coefficients $p_{k_{\alpha},\gamma}$. As the conditional state of the environment can be influenced by the experimenter, this suggests that the overall scaling parameters can be adjusted -- and hence decoherence can be suppressed -- by interfering elementary trajectories. Similarly to a single step process, one can extend the formalism to compute scaling parameters for a concatenation of $\Sys\Env$ unitaries and interventions (a detailed derivation can be found in \ref{appen:iteration}). In this case, the resulting map may \textit{not} be unital (even if all interventions are unital) but the overall scaling parameter is still meaningful for the task at hand; the smaller the scaling parameter the better the decoupling between system and environment.

Now let's revisit the example of applying the operators $\op{G}, \op{G}^2,\dots,\op{G}^d$ as in Eq.~\eqref{eqn::PermDecoup}. The scaling parameters of the resulting map for this concatenation of control operations can be calculated directly by performing a scaling-unitary decomposition of $\Phi$ (see~\ref{appen::Decom} and Ref.~\cite{FattahPreprint}). The overall scaling parameters are given by
	\begin{gather}
		\ell_{\mu\mu'}(t) = -\dfrac{1}{2}\ln\left\vert \sum_{\gamma,\gamma'}   p_\gamma p_{\gamma'}\cos\left( t \sum_{r=1}^d  \Omega_{\gamma\gamma'}\omega_{\pi_r(\mu)\pi_r(\mu')} \right)  \right\vert,
		\label{eqn::DecoupG}
	\end{gather}
where the mapping $\pi_r:\{1,\dots,d\}\longrightarrow\{1,\dots,d\},$ with $\pi_r\neq\pi_s$ for $r\neq s$, denotes a permutation of indices of the eigensubspaces of $\op{S}$ and $p_\gamma = \Tr{\op{P}^{\Env}_\gamma\rho^{\Env}_1}.$ Since the permutation is over a full cycle, the argument of the cosine function vanishes, and, as $\sum_{\gamma,\gamma'} p_{\gamma,\gamma'} = 1$, so do all of the scaling parameters. Therefore, the dynamical map $\Phi$ contains no dissipation. We have already seen an example of this decoupling in the previous section for a qubit system ($d=2$) and the shift/flip operator $\op{G} = \sigma_x$.

The fact that a concatenation of operations $\op{G},\dots,\op{G}^d$ leads to decoupling can also be derived by directly calculating the scaling parameters for interfering elementary trajectories. To this end, we express the operations $A_\alpha$ performed at $t_\alpha$ as 
\begin{gather}
\label{eqn::coeffA}
    A_{\alpha} =\sum_{\mu_{\alpha}\nu_{\alpha} , \mu'_{\alpha}\nu'_{\alpha}} g^\alpha_{\mu_{\alpha}\nu_{\alpha} ; \mu'_{\alpha}\nu'_{\alpha}} \mathcal{P}_{\mu_{\alpha}\nu_{\alpha} ; \mu'_{\alpha}\nu'_{\alpha}}\, ,
\end{gather}
where $\mathcal{P}_{\mu_{\alpha}\nu_{\alpha} ; \mu'_{\alpha}\nu'_{\alpha}}[\cdot]  = \ketbra{\mu_{\alpha}}{\nu_{\alpha}} \cdot \ketbra{\nu'_{\alpha}}{\mu'_{\alpha}}$ are elementary operators that -- just as causal breaks -- span the space of operators on the system. We emphasize that $\mathcal{P}_{\mu_{\alpha}\nu_{\alpha} ; \mu'_{\alpha}\nu'_{\alpha}}$ is not a causal break, but we choose to work in a decomposition of the form~\eqref{eqn::coeffA} for the remainder of this section to reduce notational overhead. Using this representation, the scaling parameters of the resulting map $\Phi$ for a concatenation of intermediate operations $A_\alpha$ can be calculated explicitly (see Eq.~\eqref{eq::InterScalPara}). Importantly, the resulting scaling parameters contain the coefficients $g^\alpha_{\mu_{\alpha}\mu_{\alpha-1};\mu'_{\alpha}\mu'_{\alpha-1}}$, which means that the overall scaling parameters can be tuned -- and forced to vanish -- by varying the operations $A_\alpha$ accordingly. 

For our example, one can write $A_{\alpha}=\op{G}^\alpha$ for $\alpha = 1,\ldots,d.$ The shift operator $\mathcal{G}$ can be expressed as 
    \begin{align}
        \mathcal{G}\left[\,\cdot\,\right] &= \sum_{\mu,\nu,\mu',\nu'} \delta_{\mu,\nu+1}\delta_{\mu',\nu'+1}\proj{\mu\nu;\mu'\nu'}{\,\cdot\,},\\
    \end{align}
where
    \begin{gather}
        \delta_{\mu,\nu} = \left\{ \begin{array}{ll} 1 & \mu\equiv\nu \mod d, \\ 0 & \text{otherwise.} \end{array}\right.
    \end{gather}
Similarly, the coefficients for $A_\alpha$ are of the form $g^\alpha_{\mu_{\alpha}\mu_{\alpha-1};\mu'_{\alpha}\mu'_{\alpha-1}} = \delta_{\mu_{\alpha}-\alpha,\mu_{\alpha-1}}\delta_{\mu'_{\alpha}-\alpha,\mu'_{\alpha-1}}.$ 
For $d$ interventions $A_\alpha$, the coefficient $C_\chi$ in Eq.~\eqref{eq::InterScalPara} becomes a product of delta functions which collapses the inner sum of Eq.~\eqref{eq::InterScalPara} and makes the argument of the cosine function vanish. As a result, the scaling parameter also vanishes, which reproduces the behaviour of Eq.~\eqref{eqn::DecoupG}. Additionally, Eq.~\eqref{eq::InterScalPara} allows to calculate the scaling parameters for the resulting map $\Phi$ for \textit{any} concatenation of intermediate operations $A_\alpha$.

For interactions that are not of the form $\op{H} = \op{S}\otimes\op{B},$ the dissipation can be produced both by translational as well as scaling parts. 
As already mentioned, even in this case, vanishing scaling parameters imply perfect decoupling; the overall map $\Phi$ is always CP. Consequently, a vanishing scaling part implies a vanishing translational part, as otherwise $\Phi$ would not map all states onto physical states.

\section{Conclusion}
\label{sec:conclusion}

In this article, we have constructed a notion of stochastic quantum trajectories and shown that any quantum dynamics can be expressed as a linear combination of them. However, there are quantum dynamics that cannot be expressed as a convex mixture of elementary trajectories which we call coherent processes or trajectories. This finding implies a notion of non-separability in time and is an analogue of entanglement in space. This then clearly differentiates quantum dynamics from classical dynamics. 

In the second half of the paper we show that, for non-Markovian processes, coherent control can allow for decoupling from the environment. We demonstrate this for a particular class of interaction Hamiltonians, explicitly computing the non-Markovian correlations and showing that they can indeed act as a resource for decoupling, as has been recently conjectured~\cite{arXiv:1301.2585}. Finally, we provide a geometric picture for how the quantum interference of trajectories leads to cancellation of decoherence. Our work highlights the importance for quantum control of non-Markovian dissipative dynamics.

The stochastic trajectories we consider are distinct from several other notions of trajectories in quantum mechanics. For instance, some definitions are motivated by trajectories in classical physics and used to question the foundations of quantum mechanics~\cite{bohm}. While others are more practical, and serve as a calculation tool. The most famous example is, of course, Feynman's path integral formulation of quantum mechanics~\cite{FeynmanHibbs1965} -- later extended to open evolution with a semiclassical bath~\cite{Feynman1963} -- but is also explicit in discrete time path sum representations of quantum circuits~\cite{PenneyEA2016} and stochastic path integral descriptions of continuously measured quantum systems~\cite{Chantasri2013, Chantasri2015}. The latter approach extends classical stochastic calculus~\cite{Klebaner2005} to an operationally meaningful quantum setting. Finally, the numerical \textit{quantum trajectories technique} is a central tool for many researchers working at the intersection of many-body physics and open dynamics~\cite{daly}.

The trajectory framework developed here also applies to classical probability theory, and describes stochastic evolutions with interventions. Moreover, it is clear that all classical evolutions can be expressed as a convex mixture of classical trajectories. However, these classical evolutions are more general than those without interventions and the former case fully contains the latter. Processes with interventions are used, among other things, in the field of causal modelling~\cite{pearl_causality_2009, 1367-2630-18-6-063032, oreshkov_causal_2016, allen_quantum_2017}, and in the study of $\epsilon$-transducers within the framework of computational mechanics~\cite{barnett_computational_2015, thompson_using_2017}; they are fundamentally unavoidable in the realm of quantum mechanics~\cite{BreuerEA2016, milz_kolmogorov_2017}. Introducing interventions is therefore crucial for a complete representation of \textit{quantum} stochastic trajectories.\\

\noindent\textbf{Acknowledgments.} FS is grateful to thank Sri-Trang Thong Scholarship, Faculty of Science, Mahidol University for financial support to study at its Department of Physics. FS also thanks Dr Sujin Suwanna for useful discussion and critical reading of manuscript. This work has mainly been conducted during the visit of FS to the School of Physics \& Astronomy, Monash University, Victoria, Australia. The financial support from \textit{Postgraduate Student Exchange Program 2016}, International Relations Division, Mahidol University, Thailand, is gratefully acknowledged. FS thanks the School of Physics \& Astronomy for logistic support. SM is supported by the Monash Graduate Scholarship (MGS), Monash International Postgraduate Research Scholarship (MIPRS) and the J L William Scholarship. KM is supported through Australian Research Council Future Fellowship FT160100073.

\section*{Appendices}
\appendix

\section{Unital reduced dynamics for \texorpdfstring{$\opH = \opS\otimes\opB$}{}} 
\label{app::unital}

Here, we prove that Hamiltonians of the form we are considering can only lead to unital conditional maps on the system. At time-step $t_{\alpha}$, let the total state be $\rho^{\Sys \Env}_{\alpha}$. (This state depends on the history $\mathbf{A}_{\alpha-1:1}$, but we are omitting the conditional label for simplicity). We now implement a causal break $\oneA_{l_{\alpha} k_{\alpha}}$ and then let system and environment interact to yield the state at the next time $t_{\alpha+1}$:

\begin{gather}
\rho_{t_{\alpha+1}} = p_{k_{\alpha}} \partTr{\Env}{ e^{-i \opS\otimes\opB t} \mathbf{R}_{l_{\alpha}} \otimes \rho^{\Env}_{\alpha|k_{\alpha}} e^{i \opS\otimes\opB t}} 
= p_{k_{\alpha}} \Phi_{\alpha+1:\alpha|k_{\alpha}} [\mathbf{R}_{l_{\alpha}}]\, ,
\end{gather}
where we have set $t:=t_{\alpha+1} - t_{\alpha}$. The conditional maps $\Phi_{\alpha+1:\alpha|k_{\alpha}}$ are dependent on $k_{\alpha}$ due to the dependence on $k_{\alpha}$ of the state of $\Env$ after the causal break. 

We can now expand the environment part of the Hamiltonian in its eigenbasis as $\opB = \sum_\gamma b_\gamma \opP^\Env_\gamma$ with $\opP^\Env_\gamma\opP^\Env_{\gamma^{\, \,  \prime}} =\opP^\Env_\gamma \delta_{\gamma\gamma^{\,  \prime}}$. Consequently, we have $\exp(-i\opS\otimes\opB t)= \sum_\gamma \exp(-i b_\gamma  \opS t) \otimes \opP^\Env_\gamma$. Using this, we find that the conditional maps have the form
\begin{gather}
\Phi_{\alpha+1:\alpha|k_{\alpha}} [\mathbf{R}_{l_{\alpha}}]
= \sum_{\gamma} p_{\gamma|k_{\alpha}} e^{-i b_\gamma \opS t} \ \mathbf{R}_{l_{\alpha}} \ e^{i b_\gamma \opS t}= \sum_{\gamma} p_{\gamma|k_{\alpha}} \mathbf{u}_\gamma \ \mathbf{R}_{l_{\alpha}} \ \mathbf{u}_\gamma^\dag,
\end{gather}
where $\mathbf{u}_\gamma=\exp(-i b_\gamma \opS t),$ and $p_{\gamma|k_{\alpha}}:=\Tr{ \opP^\Env_{\gamma} \, \rho^{\Env}_{\alpha|k_{\alpha}}}$ are probabilities. This is a random unitary process, the collection of which form a subset of unital processes -- those that preserve the maximally mixed states, $\Phi(\iden)=\iden$. We have proven that  all conditional maps are unital for this type of $\Sys\Env$ interaction. In what follows, we will use unitality to decompose the conditioned maps into a scaling and a rotational part.

\section{Scaling-unitary decomposition for unital dynamical maps}\label{appen::Decom}
Now we introduce a scaling-unitary decomposition of unital CPTP dynamical maps~\cite{FattahPreprint}. Let $\Hilbert$ be a $d$-dimensional Hilbert space and $\Phi$ a unital map acting on the space $\Banach(\Hilbert)$ of bounded self-adjoint operators on $\Hilbert$. As $\Banach(\Hilbert)$ can be mapped onto $\mathbb{R}^{d^2}$~\cite{bengtsson_geometry_2007}, $\Phi$ possesses a real matrix representation $\M$ acting on $\mathbb{R}^{d^2}$. In particular, using the orthonormal set of generalised Gell-Mann matrices~\cite{Kimura2003,Alicki2007} and the identity matrix
\begin{gather*}
\{\, \, f_0=\dfrac{1}{d}\iden_d,f_{\alpha}: \mathrm{Tr}(f_{\alpha} f_\beta)=\delta_{\alpha\beta}, \Tr{f_{\alpha}}=0, 1\leqslant\alpha ,\beta \leqslant d^2-1\}
\end{gather*}
as a basis of $\Banach(\Hilbert)$, the matrix elements of $\M$ are given by
	\begin{gather}
		\M_{\alpha\beta} = \mathrm{Tr}\left[f_{\alpha} \Phi\left( f_\beta\right)\right].
	\end{gather}
Any operator $\op{a}\in \Banach(\Hilbert)$ can be represented as a vector $\x(\op{a})\in \mathbb{R}^{d^2}$ with $x_{\alpha}=\mathrm{Tr}(\op{a}f_{\alpha})$ for $\alpha=0,\ldots,d^2-1.$ Unitality of $\Phi$ then translates to $\M_{\alpha 0} = \delta_{\alpha 0}$, while trace preservation implies $\M_{0\alpha} = \delta_{0\alpha}$. Consequently, $\M$ can be decomposed as $\M = 1 \oplus \tilde{\M}$; there is no transfer between $\mathrm{span}(f_0)$ and the space spanned by the traceless Gell-Mann matrices. 

Let $\varphi$ denote the isomorphism between the set of unital maps and their corresponding matrices, \textit{i.e.}, $\Phi=\iso{\M}$. We have \cite{FattahPreprint}
	\begin{gather}
	\label{eqn::DirectSum}
		\iso{\mathbf{M}} = \iso{1\oplus\tilde{\M}}= \iso{1\oplus\op{\tilde{D}}}\circ\iso{1\oplus\op{\tilde{O}}},
	\end{gather}
where $\op{\tilde{D}}$ is a positive symmetric scaling matrix, while $\op{\tilde{O}}$ is an orthogonal matrix. 
In this representation, it is easy to see that the dissipation produced by the map $\Phi,$ including the correlation between system and environment and entropy production, is contained in the scaling matrix $\op{\tilde{D}}$ alone. If, additionally, the dynamical matrix $\M$ is normal (this is satisfied for our choice of Hamiltonian), 
by an appropriate change of basis in $\R^{d^2}$, one can further simplify $\M$ to a direct sum of $2$-dimensional subspaces. This useful property fails for general unital maps (see discussion in Ref.~\cite{FattahPreprint}).

Following the procedure laid out in~\cite{FattahPreprint}, the  \textit{scaling-unitary} decomposition~\eqref{eqn::DirectSum} can be carried out explicitly for the conditioned map we are interested in (see Eq.~\eqref{eq:SubspaceEvolution}):
\begin{align}
\Phi_{\alpha+1:1|k_{\alpha},\mathbf{A}_{\alpha-1:1}}\left[\mathbf{R}_{l_{\alpha}}\right] &=	\sum_{\nu\nu'}\left(e^{-\ell^{(k_{\alpha})}_{\nu\nu'}(t)}\mathcal{P}_{\nu\nu'}\sum_{\mu\mu'}{e^{i\phi^{(k_{\alpha})}_{\mu\mu'}(t)}\proj{\mu\mu'}{\mathbf{R}_{l_{\alpha}}}}\right)\label{eq:naivemap}\\
&= \left( \iso{\op{\tilde{D}}^{(k_{\alpha})}} \circ \iso{\op{\tilde{O}}^{(k_{\alpha})}} \right) \left[\mathbf{R}_{l_{\alpha}} \right] \, ,
\end{align}	
where 
\begin{align}
& \ell^{(k_{\alpha})}_{\mu\mu'}(t)=-\dfrac{1}{2}\ln\displaystyle\left\vert  \sum_{\gamma,\gamma'}   p_{k_{\alpha},\gamma} p_{k_{\alpha},\gamma'}\cos\left( \Omega_{\gamma\gamma'}\omega_{\mu\mu'}t  \right)\right\vert, \\ 
& \phi^{(k_{\alpha})}_{\mu\mu'}(t) = \arctan\left(\dfrac{\sum_{\gamma}   p_{k_{\alpha},\gamma} \sin\left( b_{\gamma}\omega_{\mu\mu'}t \right) }{\sum_{\gamma}  p_{k_{\alpha},\gamma}\cos\left( b_{\gamma}\omega_{\mu\mu'}t \right)} \right),
\end{align}
and $\Omega_{\gamma\gamma'} = b_\gamma-b_{\gamma'}.$  Note that the operations $\proj{\mu\mu'}{\cdot}$ are orthogonal in both indices. From this decomposition, it is clear that the dissipation, measured as purity of the final state, appears only in the scaling matrix $\op{\tilde{D}}^{(k_{\alpha})}$ as mentioned. Furthermore, it can be also seen that the history dependency affects both scaling and rotation parts of the dynamical matrices. 

\section{Scaling parameters for quantum trajectories}\label{appen:iteration}

In a similar fashion to the single intervention case, one can derive the scaling part for the resulting map for a sequence of interventions. 
To this end, for notational convenience, we expand the causal breaks $\oneA_{l_{\alpha} k_{\alpha}}$ at time step $t_{\alpha}$ in the basis of the system part of the Hamiltonian as 
	\begin{gather}
		\oneA_{l_{\alpha} k_{\alpha}}\left[\,\cdot\, \right] = \displaystyle\sum_{\mu_{\alpha}\nu_{\alpha} ; \mu'_{\alpha}\nu'_{\alpha}}f_{\mu_{\alpha}\nu_{\alpha} ; \mu'_{\alpha}\nu'_{\alpha}}^{(l_{\alpha} k_{\alpha})}\proj{\mu_{\alpha}\nu_{\alpha} ; \mu'_{\alpha}\nu'_{\alpha}}{\,\cdot\,}
		\label{eq::InterHamilBase}
	\end{gather}
where $\proj{\mu_{\alpha}\nu_{\alpha} ; \mu'_{\alpha}\nu'_{\alpha}}{\,\cdot\,} = \vert\mu_{\alpha}\rangle\langle\nu_{\alpha}\vert\cdot\vert\nu'_{\alpha}\rangle\langle\mu'_{\alpha}\vert$ is a subspace transfer operator with respect to the operator $\op{S}$ and $f_{\mu_{\alpha}\nu_{\alpha} ; \mu'_{\alpha}\nu'_{\alpha}}^{(l_{\alpha} k_{\alpha})} = \mathrm{Tr}\left\{\vert\mu_{\alpha}\rangle\langle\mu'_{\alpha}\vert\oneA_{l_{\alpha} k_{\alpha}}\left[\,\vert\nu_{\alpha}\rangle\langle\nu'_{\alpha}\vert\, \right]\right\}.$ As above, the subscript $\alpha$ refers to the time step to which $\mathcal{P}_{\mu_{\alpha}\nu_{\alpha} ; \mu'_{\alpha}\nu'_{\alpha}}$ corresponds. 
Representing an arbitrary operation $A_{\alpha}=\displaystyle\sum_{l_{\alpha},k_{\alpha}}a_{l_{\alpha} k_{\alpha}}\oneA_{l_{\alpha} k_{\alpha}}$ in terms of causal breaks, we obtain
\begin{gather}
A_{\alpha} = \displaystyle\sum_{\mu_{\alpha}\nu_{\alpha} , \mu'_{\alpha}\nu'_{\alpha}} g^\alpha_{\mu_{\alpha}\nu_{\alpha} ; \mu'_{\alpha}\nu'_{\alpha}} \mathcal{P}_{\mu_{\alpha}\nu_{\alpha} ; \mu'_{\alpha}\nu'_{\alpha}}
\end{gather}
where $g^\alpha_{\mu_{\alpha}\nu_{\alpha} ; \mu'_{\alpha}\nu'_{\alpha}} = \displaystyle\sum_{l_{\alpha},k_{\alpha}}f_{\mu_{\alpha}\nu_{\alpha} ; \mu'_{\alpha}\nu'_{\alpha}}^{(l_{\alpha} k_{\alpha})}a_{l_{\alpha} k_{\alpha}}.$

From Eq.~\eqref{eqn::TrajSum}, we know that the overall reduced dynamics for an elementary trajectory (with initial $\Sys\Env$ state $\rho_1^{\Sys\Env}=\rho_1\otimes\rho^\Env_1$)  is given by a CP map $\Phi^{(\xi)}$ that depends on the respective trajectory. Consequently, the overall dynamics for a sequence $\mathbf{A}_{N-1:1}$ can be written as a linear combination of elementary trajectories, \textit{i.e.}, $\rho_N(\mathbf{A}_{N-1:1}) = \sum_{\xi\in\pk}a^{(\xi)}\Phi^{(\xi)}\left(\rho_1 \right)$. As for the case of a single causal break, the simple form of the $\Sys\Env$ Hamiltonian allows one to explicitly calculate the (trajectory dependent) maps $\Phi^{(\xi)}$ for an elementary trajectory:
	\begin{gather}
		\Phi^{(\xi)}\left[\rho_1\right] = \sum_{\stackrel{\mu_{N-1},\nu_1}{\mu'_{N-1},\nu'_1}}\left[\sum_{\vec{\mu}_{N-2},\vec{\mu}'_{N-2}}\sum_\gamma C_{\stackrel{\gamma,\vec{\mu}_{N-1},\nu_1}{~~\vec{\mu}'_{N-1},\nu'_1}}^{(\xi)} e^{-i b_\gamma \omega_{\vec{\mu}_{N-1},\vec{\mu}'_{N-1}} t}\right] \proj{\mu_{N-1}\nu_1 ; \mu'_{N-1}\nu'_1}{\rho_1},
		\label{eq::TrajDepMap}
	\end{gather}
where 
\begin{gather}
\omega_{\vec{\mu}_{N-1},\vec{\mu}'_{N-1}}=(1/{N})\displaystyle\sum_{\alpha=1}^{N-1} \omega_{\mu_{\alpha}\mu'_{\alpha}}
\end{gather}
and 
\begin{gather}
C_{\stackrel{\gamma,\vec{\mu}_{N-1},\nu_1}{~~\vec{\mu}'_{N-1},\nu'_1}}^{(\xi)} = \mathrm{Tr}\left(\opP^\Env_\gamma\rho_1^\Env\right)\left(f^{(l_1(\xi)k_1(\xi))}_{\mu_1\nu_1 ; \mu'_1\nu'_1}\displaystyle\prod_{\alpha=2}^{N-1} f^{(l_{\alpha}(\xi)k_{\alpha}(\xi))}_{\mu_{\alpha}\mu_{\alpha-1} ; \mu'_{\alpha}\mu'_{\alpha-1}}\right) := C_{\gamma,N-1}^{(\xi)}\, , 
\end{gather}
 
where we employ the shorthand notation $\vec{\mu}_M = \left(\mu_{M},\mu_{M-1},\ldots,\mu_1\right)$.
The coefficients $C_{\gamma,N-1}^{(\xi)}$ contain the choices of causal breaks for the respective elementary trajectory $\xi$.  Consequently, the reduced dynamical maps $\Phi^{(\xi)}$ are generally trajectory dependent. 

Now, we can interfere elementary trajectories to obtain the overall map $\Phi$ for a sequence of CPTP maps $A_\alpha$ and calculate its corresponding scaling parameters. To this end, we first rewrite $\Phi = \sum_\xi a^{(\xi)}\Phi^{(\xi)}$ differently as 
   \begin{align}
        \Phi\left[\rho_1\right] &= \sum_{\xi\in\pk}a^{(\xi)}\sum_{\stackrel{\vec{\mu}_{N-1},\nu_1}{\vec{\mu}'_{N-1},\nu'_1}}\left[\sum_\gamma C_{\stackrel{\gamma,\vec{\mu}_{N-1},\nu_1}{~~\vec{\mu}'_{N-1},\nu'_1}}^{(\xi)} e^{-i b_\gamma \omega_{N-1:1} t}\right] \proj{\mu_{N-1}\nu_1 ; \mu'_{N-1}\nu'_1}{\rho_1 } \\
        &= \sum_{\stackrel{\vec{\mu}_{N-1},\nu_1}{\vec{\mu}'_{N-1},\nu'_1}}\left[\sum_\gamma C_{\stackrel{\gamma,\vec{\mu}_{N-1},\nu_1}{~~\vec{\mu}'_{N-1},\nu'_1}} e^{-i b_\gamma \omega_{N-1:1} t}\right] \proj{\mu_{N-1}\nu_1 ; \mu'_{N-1}\nu'_1}{\rho_1 },
        \label{eqn::overall_dynamics}
    \end{align}   
where 
    \begin{align}
        C_{\stackrel{\gamma,\vec{\mu}_{N-1},\nu_1}{~~\vec{\mu}'_{N-1},\nu'_1}} &= \sum_{\xi\in\pk}a^{(\xi)}C_{\stackrel{\gamma,\vec{\mu}_{N-1},\nu_1}{~~\vec{\mu}'_{N-1},\nu'_1}}^{(\xi)}
            = \mathrm{Tr}\left(\opP^\Env_\gamma\rho_1^\Env\right)\left(\prod_{\alpha=0}^{N-1}g^\alpha_{\mu_{\alpha}\mu_{\alpha-1};\mu'_{\alpha}\mu'_{\alpha-1}} \right),
        \label{eq::CoeffControl}
    \end{align}
and $\mu_0=\nu_1,$  $\mu'_0=\nu'_1.$ 

We emphasize that the map $\Phi$ in Eq.~\eqref{eqn::overall_dynamics} is in general not unital, even if all the control operations are unital and the $\Sys\Env$ Hamiltonian is of the form $\mathbf{S}\otimes \opB$. While this is the case for Markovian dynamics, it fails to generally apply in the non-Markovian case. Non-unitality makes it harder to isolate the decoherence part of $\Phi$; besides the rotational and the scaling part, a non-unital map also displays a translational part that contains information about decoherence. As a vanishing scaling part implies unitary dynamics, the objective of control could nonetheless be to minimize the scaling part of $\Phi$. Consequently, we perform the scaling-unitary decomposition of the interfered map $\Phi$ and obtain scaling parameters of the form
    \begin{gather}
        \ell_{\stackrel{\mu_{N-1},\nu_1}{\mu'_{N-1},\nu'_1}}(t)=-\dfrac{1}{2}\ln\displaystyle\left\vert  \sum_{\stackrel{\vec{\mu}_{N-2},\vec{\mu}''_{N-2}}{\vec{\mu}'_{N-2},\vec{\mu}'''_{N-2}}}\sum_{\gamma,\gamma'}  C_{\chi} \cos\left( \left(\Omega_{\gamma\gamma'}\omega_{\mu_{N-1},\mu'_{N-1}} + b_\gamma\omega_{\vec{\mu}_{N-2},\vec{\mu}''_{N-2}}-b_{\gamma'}\omega_{\vec{\mu}'_{N-2},\vec{\mu}'''_{N-2}}\right) t  \right)\right\vert,
        \label{eq::InterScalPara}
    \end{gather}
where
	\begin{gather}
		C_{\chi} := \left(g^\alpha_{\mu_{N-1}\mu_{N-2} ; \mu'_{N-1}\mu''_{N-2}}g^{*}_{\mu_{N-1}\mu'_{N-2} ; \mu'_{N-1}\mu'''_{N-2}}\right)C_{\stackrel{\gamma,\vec{\mu}_{N-2},\nu_1}{~~\vec{\mu}''_{N-2},\nu'_1}}C^{*}_{\stackrel{\gamma',\vec{\mu}'_{N-2},\nu_1}{~~\vec{\mu}'''_{N-2},\nu'_1}},
		\label{eq::C_factor}
	\end{gather}
and $\chi = \{\mu_{N-1},\nu_1,\mu'_{N-1},\nu'_1, \vec{\mu}_{N-2},\vec{\mu}''_{N-2},\vec{\mu}'_{N-2},\vec{\mu}'''_{N-2},\gamma,\gamma' \}$ denotes all indices of the summation.

The coefficients $C_{\gamma,N-1}$ explicitly depend on the choice of operations $A_\alpha$. Thus, the scaling parameters $\ell_{\stackrel{\vec{\mu}_{N-1},\nu_1}{\vec{\mu}'_{N-1},\nu'_1}}(t)$ -- and consequently, the coherence of the overall map $\Phi$ -- can be controled by adjusting the intervention parameters $g^\alpha_{\mu_{\alpha}\mu_{\alpha-1};\mu'_{\alpha}\mu'_{\alpha-1}}$ in Eq.~\eqref{eq::CoeffControl}. Due to the availability of entanglement preserving interventions, this degree of control is strictly greater in quantum mechanics than it would be classically.

\newcommand{\newblock}{}
\bibliographystyle{apsrev4-1}
\bibliography{extraction.bib}
\end{document}